\documentclass[10pt,singlecolumn,letterpaper]{article}
\usepackage{times}
\usepackage{graphicx} 
\usepackage{subfigure} 
\usepackage{amsmath}
\usepackage{epstopdf}
\usepackage{float}
\usepackage{courier}
\usepackage{cite}

\usepackage{dblfloatfix}

\usepackage[utf8]{inputenc}
\usepackage[english]{babel}
 
\usepackage{algorithm}
\usepackage{algorithmic}
\usepackage{listings}
\usepackage{color}

\usepackage{hyperref}

\usepackage{iccv}
\iccvfinalcopy
\usepackage[section]{placeins}


\begin{document}

\title{\textbf{Pathology Extraction from Chest X-Ray Radiology Reports: A Performance Study}}

\author{Tahsin Mostafiz$^1$, Khalid Ashraf$^2$\\
$^1$Semion, House 167, Road 3, Mohakhali DOHS, Dhaka, Bangladesh.\\
$^2$Semion, 5 Newell Rd., St 12,  Palo Alto, CA 94303, USA.\\
{\tt\small tahsinmostafiz314@gmail.com, khalid@semion.ai }
}

\maketitle




\begin{abstract}

Extraction of relevant pathological terms from radiology reports is important for correct image label generation and disease population studies. In this letter we compare the performance of some known application program interface (APIs) for the task of thoracic abnormality extraction from radiology reports. We explored several medical domain specific annotation tools like Medical Text Indexer(MTI) with Non-MEDLINE and Mesh On Demand(MOD) options and generic Natural Language Understanding (NLU) API provided by the IBM cloud. Our results show that although MTI and MOD are intended for extracting medical terms, their performance is worst compared to generic extraction API like IBM NLU. Finally, we trained a DNN-based Named Entity Recognition (NER) model to extract the key concept words from radiology reports. Our model outperforms the medical specific and generic API performance by a large margin. Our results demonstrate the inadequacy of generic APIs for pathology extraction task and establishes the importance of domain specific model training for improved results. We hope that these results motivate the research community to release larger de-identified radiology reports corpus for building high accuracy machine learning models for the important task of pathology extraction.


\end{abstract}

\section{Introduction}
\label{intro}
High quality deep learning models require large data set with high quality labels for training. In the medical domain, the data often exists in hospital electronic health records(EHR) and PACS systems. Building corpus for various task requires automated tools to extract correct labels from the vast amount of data. For example, for developing corpus for radiology image to pathology mapping, correct pathology labels need to be extracted from radiology reports using automated annotation tool. The extraction step is important for many other tasks like disease population analysis, findings to clinical action analysis etc. Low accuracy pathology extraction tools can result in high percentage of wrong labels on large radiology corpus and subsequently result in low accuracy machine learning models developed using these corpus \cite{wang2017chestx,rajpurkar2017chexnet,Quicktho72:online}. In general, the difficulties of pathology extraction are ruling out unnecessary words, phrases from a report and finding the relevant terms that best expresses the findings and impression of that report. Further difficulties arise in medical term extraction since these terms are very rare in natural language and hence generic APIs fare poorly with these terms. Some APIs have been developed for medical specific entity extraction and annotation tasks. The US National Library of Medicine (NLM) established the Indexing Initiative project in 1996 ~\cite{aronson2012nlm}. This project developed The NLM Medical Text Indexer (MTI) and it has been functional and providing automated indexing recommendations since 2002\cite{aronson2012nlm}. Gobbel \textit{et al.}\cite{gobbel2014assisted} also developed a tool named RapTAT, to assist in annotation. The tool annotates probable phrases of interest within a report in iteration and provides the annotations to a reviewer for correction. Using the corrected annotations the system updates its machine learning-based model. RapTAT was used for extracting concepts related to quality of care during treatment of heart failure. The annotators reviewed total of 404 clinical notes either manually or using RapTAT. Tonin \textit{et al.}~\cite{tonin2017annotating} developed a machine learning based text annotator which extracts mentions or indications of coronary artery disease in unstructured clinical reports. The performance of publicly available medical specific APIs for pathology extraction is mostly unknown or non-standardized since they are reported on private data sets or in combination with human annotators. We felt the need to benchmark the performance of radiology report annotation using the public API on a public data set so that different approaches can be compared and future progress can be benchmarked. Thus in this paper, we analyze the pathology extraction performance of several publicly available APIs (medical and non-medical) on the publicly available Indiana data radiology report data set. We also show the performance of our deep learning based NER model that outperforms the existing public APIs by a large margin. This is an early work on a relatively small radiology report corpus. Previously we had reported the first x-ray image classification and localization benchmark results for 20 different pathologies on this dataset\cite{islam2017abnormality}. We hope that these results will encourage the research community to build and release larger radiology report corpus to study this important task of entity extraction from radiology reports. Our contributions in this work are:
\begin{itemize}
    \item First report of pathology extraction from chest X-Ray radiology reports. 
    \item We present the performance and limitations of the two medical specific annotation tools and one generic API i.e. IBM NLU.
    \item We show that a deep neural network based architecture trained on the task specific data outperforms the generic tools by a large margin. 
    \item Our results emphasize the importance of task specific ML model training and data set development. 
\end{itemize}

This paper is organized as follows. In section 2, we review the related works, in section 3, we describe the task to motivate the work. The experiment with the pre-processing steps, evaluation metrics and the methods are briefly  described in section 4. In section 5, we present our results and discuss some qualitative features of the results in section 6. Finally we conclude in section 7. The supplementary materials section contains sample reports and the annotations performed by MTI tools, NLU, and our NER model.

\section{Related Works}
\label{relWork}
Demner-Fushman \textit{et al.} \cite{demner2016preparing} explored automatic and manual approaches to annotation, as well as developed a small controlled vocabulary of chest x-ray indexing terms and guidelines for manual annotation of radiology reports. They used 3,955 de-identified chest radiology reports from the Indiana chest X-Ray dataset \cite{DBLP:journals/jamia/Demner-FushmanK16}. First, they annotated the reports with two annotators, both trained in medical informatics and experienced in medical document annotation. Then they used MTI (A Medical Text Indexer that assigns MeSH terms) and SGindexer that uses MetaMap to extract asserted Unified Medical Language System(UMLS) \cite{humphreys1989building,UMLSTerm42:online}  concepts in the Disorders and Procedures semantic groups to annotate over the same report and compared their performances. 

For evaluation of the annotation tool with manually reviewed reports as ground truth, the annotators divided each annotation into the following classes: correct, neutral, somewhat incorrect, and incorrect. Annotations were judged correct if a major finding was correctly identified. They were judged neutral if the annotation was correct, but the term described trivial anatomy or findings. Annotations were judged somewhat incorrect if a part of the term that was captured did not have an appropriate sense in any of the source vocabularies. Annotations were judged incorrect if automatic annotation captured a term that was negated or was not stated in the report. The neutral terms were ignored during computing precision and recall and the somewhat incorrect and incorrect terms were judged to be false positives. The annotators assigned one of the aforementioned 4 labels to annotation done by both tools on each report and found that MTI has a recall of 28.7\%  and precision of 28.9\%. On the other hand, SGindexer has a precision of 73.3\% and recall of 40.5\%. 

Hassanzadeh \textit{et al.}\cite{hassanzadeh2016evaluation} compared several annotation tools on electronic health reports. They used \textit{ShARe/CLEF task corpus}\cite{ShAReCLE61:online} as gold standard data where the correct span of text which reflect the concept of the reports was identified by human experts. They experimented with \textit{MetaMap}, \textit{NCBO Annotator}\cite{jonquet2009open}, \textit{QuickUMLS}\cite{soldaini2016quickumls} and \textit{Ontoserver}\cite{mcbride2012using} tools, and studied the performance of  identification of correct text spans. For evaluation purpose, they counted an outcome to be true positive (TP) if a system identified a disorder in the same span as that of the gold standard. False positive (FP) was defined as the identification of an incorrect span, and their algorithm triggered false negative (FN) if a system could not identify a disorder-span that was identified by the expert assessors. Their precision was 0.8076, 0.7679. 0.9058 and 0.8008 using \textit{MetaMap, NCBO~, Ontoserver} and \textit{QuickUMLS}. They obtained recall scores of 0.6695, 0.3758, 0.6292 and 0.6893 respectively on this task.

Mirhosseini \textit{et al.}~\cite{mirhosseini2014medical}  applied a subset of the train set of \textit{ShARe/CLEF} data to compare MetaMap, Ontoserver and several numbers of standard IR techniques for concept recognition. They evaluated the performance in identifying \textit{Concept Unique Identifier} (CUI) terms by measuring \textit{Reciprocal Rate} (RR) and \textit{success@k}, which measures whether a relevant report has been retrieved up to a cut-off k (k = 1, 5, 10). They found the RR values for \textit{MetaMap} and \textit{Ontoserver} are 0.2723 and 0.6166.\\ 


\section{Tasks}
\label{taskDescription}
\begin{figure*}[!h]
\begin{center}
\includegraphics[width=0.8\textwidth, 
height=.15\textheight]{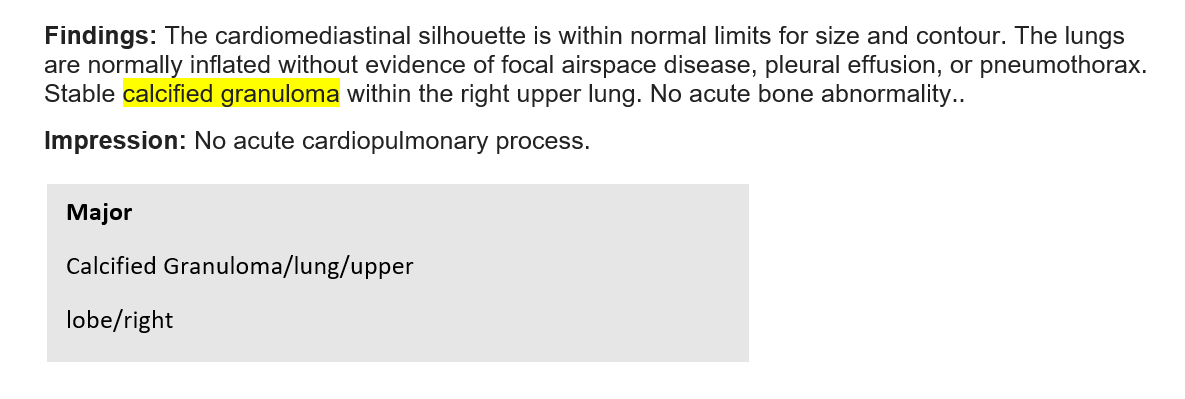}
\caption{An example of a radiology report from Indiana dataset.}
\label{fig:example_report}
\end{center}
\end{figure*}

\textit{Findings} and \textit{Comparison} sections of radiology reports contain vital information. Typically relevant pathological terms are found in these two sections. The highlighted words in the report shown in Figure \ref{fig:example_report}, taken from Indiana dataset, indicate the relevant pathological terms.
The report contains multiple pathological terms such as  \textit{cardiomediastinal silhouette, focal airspace disease, pleural effusion, pneumothorax, calcified granuloma.} All of these terms except for \textit{calcified granuloma} are negated. The task of an annotation tool would be to extract the \textit{calcified granuloma} term only. A typical pathological word search would result in extracting other negated terms as well. This is why we need a dedicated tool for extracting relevant terms.

Pathology label extraction on the Inidiana chest X-Ray dataset have been performed by Demner-Fushman \textit{et al}. \cite{demner2015annotation}. They annotated 3955 de-identified chest radiology reports with the help of two expert radiologists, both trained in medical informatics and experienced in medical document annotation. A small controlled vocabulary of chest x-ray indexing terms and guidelines was created for annotation task. Using this vocabulary, the radiologists independently annotated each report. A comparison program was implemented to find disagreements between the annotators. The program compared annotations for each document and identified missing terms and attributes. The output of the program indicated which terms and attributes were missing for each annotator for a given report. The annotators then reconciled the disagreements. They used these annotations as ground truth. Then they used MTI and SGindexer that uses MetaMap to extract asserted Unified Medical Language System(UMLS) concepts in the Disorders and Procedures semantic groups to annotate over the same report and compared their performances. We've also used annotations provided by Demner-Fushman \textit{et al.}\cite{DBLP:journals/jamia/Demner-FushmanK16} to our dataset and MeSH \cite{MedicalS14:online} terms available for each report.

\section{Experiments}
\label{experiments}
\subsection{Dataset Preprocessing}
For our annotation task, we considered only \textit{Findings} and \textit{Impression} sections of the reports. We’ve used the provided MeSH terms as our ground truth annotations. For indexing with NLM tools, we assigned a dummy PMID code with each modified report. For annotation using NLU, we tokenized each sentence from \textit{Findings} and \textit{Impression} for each report and joined them. For training the NER model, we used the tokenized sentences created for NLU and assigned each word with either \textit{KEYWORD} or \textit{NON-KEYWORD} term. If a particular word/phrase is present in both ground truth text and a report, the word is tagged with \textit{KEYWORD} for that report. Otherwise, the word is tagged \textit{NONKEYWORD}.

\subsection{Evaluation Metrics}
Indiana dataset contains over 3,955 annotated radiology chest x-ray reports and corresponding reports are publicly available. The reports are formatted according to RSNA standards\cite{ChestXra42:online}. These reports are de-identified and manually annotated MeSH terms are provided for each report. For our experiment, we considered these data as gold standard and used these annotations as ground truth.  

\subsection{Methods}
\begin{itemize}
\item \textbf{Web based indexing tools and APIs provided by U.S. National Library of Medicine (NLM) for annotation} We used the following options:
 
 \begin{itemize}
     \item  Batch MTI indexer tool with Mesh On Demand as filtering option: MeSH is a comprehensive controlled vocabulary created by indexing relevant articles and books. Mesh On Demand tool extracts words associated with medical terminology from a report. We submitted ‘Findings’ and ‘Impression’ section of each report with a dummy PMID number and requested for batch indexing. 
 
    \item MTI annotated with Non-MEDLINE option with Default for Non-MEDLINE Text as Pre-package filtering option: MEDLINE is a bibliographic database of life sciences and biomedical information \cite{wiki:xxx}. Annotation using MEDLINE requires PMID code which is available for journals and reports deposited in PubMed Central (PMC). Since we did not have access to that information, we annotated our reports using Non-MEDLINE option.    
 \end{itemize}

\item \textbf{Natural Language Understanding Service provided by IBM}: IBM NLU tool \cite{NaturalL39:online} can analyze and find out the following key concepts from a given text: Concept, Category, Emotion, Entities, Keywords, Relations, Semantic Roles and Sentiments. We used a python library that allows a user to send a request for text analysis using their API and credentials. We used ‘Entity’ to extract keywords from reports. Each ‘Entity’ word is assigned with a  ‘Sentiment’ score. A negative sentiment score for an ‘Entity’ triggered word means that particular word is negated. Each word with a positive ‘sentiment’ score is extracted. The extracted words for each report were joined to form a sentence. Each sentence was considered as an annotation for a report.

\item \textbf{Name Entity Recognition (NER)}: The Name Entity Recognition algorithms identifies name entities(i.e., Person, location or organization) from a given text. We used the method described by Chiu \textit{et al.}\cite{chiu2015named} and used it to extract key concepts from a report. We split the data set into 80:10:10 ratio for train, validation and test set. For training, we used glove 100d word embedding and trained our network for 100 epochs.
\end{itemize}

\section{Results}
\label{results}
For our task, we word tokenized each annotation word (both prediction and actual) for each sentence and formed a sentence like structure by joining each word for annotation with 
space and adding full stop at the end. For example: the annotation \textit{Aorta,   Thoracic,Cicatrix, Costophrenic Angle, Thickening} is converted into \textit{Aorta thoracic cicatrix costophrenic Angle thickening}.If there was no predicted annotation available for a report, we assumed the prediction would be \textit{normal}. Table I is calculated for the following pathology terms : opacity, aorta, fractures, osteophyte, scoliosis, density, pneumothorax, cardiomegaly, emphysema, arthritis, granuloma, kyphosis, pneumonia, spondylosis, deformity, hypertension, consolidation, mass, thickening, hernia, lucency, consolidation, bronchiectasis.

\begin{table*}[!h] 
\caption{BLEU scores calculated for different annotation tools and our model}
\label{BLEU}
\vskip 0.15in
\begin{center}
\begin{small}
\begin{sc}
\begin{tabular}{lccccr}
\hline
 & Batch MTI indexer   & MTI annotated   & IBM NLU (\%) & NER (\%) \\ & tool with Mesh On Demand (\%) &  with Non-MEDLINE (\%) & & [Ours]\\
\hline
BLEU-1    & $23.52\%$ 	& $24.41\%$  & $45.05\%$  & $\bf49.53\%$\\
BLEU-2      & $1.23\%$ & $1.24\%$  	& $3.96\%$ 	& $\bf4.82\%$ \\
BLEU-3      & $0.11\%$ 	& $0.11\%$  	& $0.39\%$ 	& $\bf0.48\%$ \\
BLEU-4   & $0.01\%$ 	& $0.011\%$  	& $0.04\%$ 	& $\bf0.05\%$\\
\hline
\end{tabular}
\end{sc}
\end{small}
\end{center}
\end{table*}

The performances of each tool were evaluated based on their BLEU(\textit{Bilingual Evaluation Understudy Score}) scores~\cite{papineni2002bleu}. BLEU score is typically calculated for evaluating text generation for a specific natural language processing task. BLEU score compares a newly generated sentences to a reference sentence. A perfect score of 1.0 means that the generated sentence exactly matches the reference sentence whereas a score of 0.0 means there is no match between them. We used the actual summary sentences as reference and predicted sentences as candidate. Table \ref{BLEU} shows the BLEU scores calculated for each tool and method. We also calculated precision, recall and F1 scores for each tool shown in Table \ref{overall_f}. 


We further calculate the precision, recall and F1 scores for individual abnormality like cardiomegaly and opacity. The results are summarized in Table \ref{opacity_f} (opacity) and Table \ref{cardiomegaly_f} (cardiomagely). We find that for individual pathology extraction task, the MTI annotation tool with NON-MEDLINE option and MESH ON DEMAND show no extraction capability at all and hence result in 0 precision-recall for \textit{opacity} term extraction. IBM NLU tool performs reasonably better than the MTI tools. Our NER tool trained on chest X-Ray reports perform significantly better than all the other tools. For the cardiomegaly extraction task, the MTI tools perform better than our NER or IBM's NLU interface.

\begin{table*}[!h]
\caption{Precision, Recall, F1 scores calculated for different annotation tools and our model}
\label{overall_f}
\vskip 0.15in
\begin{center}
\begin{small}
\begin{sc}
\begin{tabular}{lccccr}
\hline
 & Precision(\%)   & Recall (\%) & F1 Score (\%) \\
\hline
MTI annotated with Non-MEDLINE option 	& $6.25\%$  & $29.41\%$  & $10.30\%$\\
Batch MTI indexer tool with Mesh On Demand & $11.38\%$  	& $23.89\%$ 	& $15.42\%$ \\
NLU   	& $21.46\%$  	& $34.55\%$ 	& $26.47\%$ \\
NER [Ours]   & $\bf45.34\%$  	& $\bf55.51\%$ 	& $\bf49.91\%$\\

\hline
\end{tabular}
\end{sc}
\end{small}
\end{center}
\vskip -0.1in
\end{table*}

\begin{table*}[!h]
\caption{Precision, Recall, F1 scores calculated for different annotation tools and our model for extraction of OPACITY.}
\label{opacity_f}
\vskip 0.15in
\begin{center}
\begin{small}
\begin{sc}
\begin{tabular}{lccccr}
\hline
 & Precision(\%)   & Recall (\%) & F1 Score (\%) \\ \\
\hline

MTI annotated with 	& $0.00\%$  & $0.00\%$  & $0.00\%$\\
Non-MEDLINE option  
\\
Batch MTI indexer  & $0.00\%$  	& $0.00\%$ 	& $0.00\%$ \\
tool with Mesh On Demand 
\\
NLU   	& $16.66\%$  	& $10.14\%$ 	& $12.61\%$ \\ \\
NER [Ours]   & $\bf92.85\%$  	& $\bf74.28\%$ 	& $\bf82.53\%$\\

\hline
\end{tabular}
\end{sc}
\end{small}
\end{center}
\vskip -0.1in
\end{table*}

\begin{table*}[h]
\caption{Precision, Recall, F1 scores calculated for different annotation tools and our model for extraction of CARDIOMEGALY.}
\label{cardiomegaly_f}
\vskip 0.15in
\begin{center}
\begin{small}
\begin{sc}
\begin{tabular}{lccccr}
\hline
 & Precision(\%)   & Recall (\%) & F1 Score (\%) \\ \\
\hline

MTI annotated with 	& $56.81\%$  & $\bf71.42\%$  & $63.29\%$\\
Non-MEDLINE option  
\\
Batch MTI indexer  & $66.66\%$  	& $62.85\%$ 	& $\bf64.70\%$ \\
tool with Mesh On Demand 
\\
NLU   	& $11.11\%$  	& $2.85\%$ 	& $4.54\%$ \\ \\
NER [Ours]   & $\bf90.90\%$  	& $28.57\%$ 	& $43.47\%$\\

\hline
\end{tabular}
\end{sc}
\end{small}
\end{center}
\vskip -0.1in
\end{table*}

\section{Discussion}
\label{discussion}
The DNN-based NER method that we trained on the task specific radiology reports performed significantly better than other tools in extracting pathological terms. We achieved better \textit{BLEU} scores and F1 scores with this method. It also learned the sentence patterns where a particular term is negated. Besides, as we used word embedding to train our model, it could differentiate between similar and no-similar words. The relatively higher accuracy of our model results from training the model for the specific task of chest X-Ray pathology extraction. 
    
\begin{figure*}[!ht]
\begin{center}
\includegraphics[width=1.0\textwidth, 
height=.25\textheight]{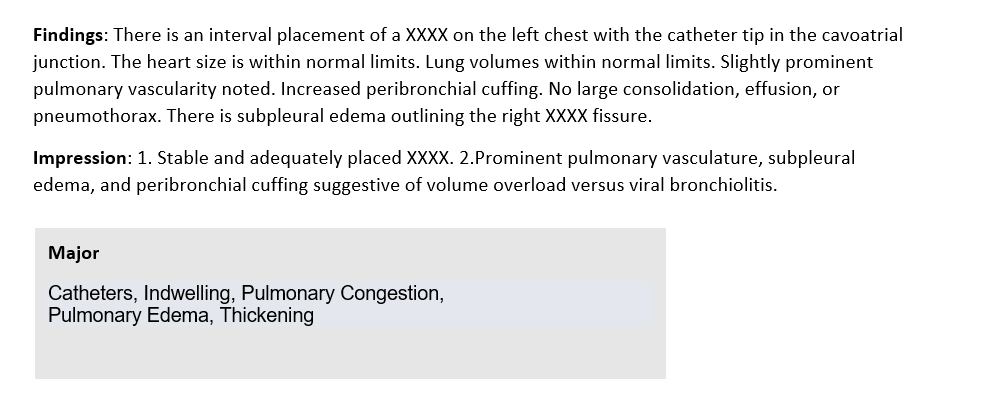}
\caption{An example of a radiology report from Indiana dataset.}
\label{fig:example_report_2}
\end{center}
\end{figure*}

MTI tools are mainly designed for the explanation of pathological terms present in a report. They triggered a lot of unnecessary words/terms/phrases which are not present in the actual report. For example, Batch MTI indexer tool with Mesh On Demand detected the term ‘CABG’ as some sort of chemical compound and annotated with ‘nitro compounds hydrogen-ion’. As a result, the summary text for radiology reports in the Indiana data set sometimes contain terms that are not present in the actual report. For example: the ground truth annotation for the report in Figure \ref{fig:example_report_2} is: \textit{Catheters, Indwelling, Pulmonary Congestion, Pulmonary Edema, Thickening}. Among these the terms \textit{Indwelling, Pulmonary Congestion and Thickening} are absent in the actual report. These are some of the anomalies that future upgrades to the MTI api will need to incorporate to improve performance.

\section{Conclusion}
\label{conclusion}
In summary, we explored several annotation tools for chest X-Ray pathology extraction in this letter. We found that the performance of the existing annotation tools are not satisfactory for the pathology extraction task. NLU is better at detecting negation but couldn’t identify radiology specific terms with positive sentiment score most of the time. Our chest X-Ray report trained NER model gave significantly better performance compared to the generic IBM NLU and medical specific MTI APIs. However, there is still opportunity for significant improvement if the model is trained with relatively large radiology report corpus. 

High accuracy pathology extraction is of utmost importance for building highly accurate machine learning models using large radiology corpus. Inferior performance in this step affects all the subsequent processing steps and an overall poor machine learning model. In this paper, we point out the inadequacy of the existing pathology extraction tools and demonstrated improved performance with our DNN-based NER tool trained on chest X-Ray reports. We hope that these results will motivate building large de-identified radiology report corpus for training accurate extraction models.  

\section{Acknowledgement}
\label{acknowledge}
This research used resources of the National Energy Research Scientific Computing Center, a DOE Office of Science User Facility supported by the Office of Science of the U.S. Department of Energy under Contract No. DE-AC02-05CH11231. Thanks to Prabhat at NERSC for sharing his allocation on the NERSC computers.

\bibliography{refs}
\bibliographystyle{IEEEbib}

\clearpage

\section{Supplementary Materials}~\label{supplimentaryMaterials}
This section contains some examples of chest X-ray reports, relevant pathological terms, terms extracted by MTI tools, NLU and NER.  
\begin{enumerate}
    \item \textbf{Findings}: \textit{Moderate cardiomegaly. Mild bilateral costophrenic XXXX blunting and fissural thickening, interstitial opacities greatest in the central lungs and bases with indistinct vascular margination. Dense right lower lobe nodule and right hilar calcifications suggest a previous granulomatous process.} \\ 
    \textbf{Impression}:  \textit{1. Cardiomegaly and small bilateral pleural effusions 2. Abnormal pulmonary opacities most suggestive of pulmonary edema, primary differential diagnosis atypical infection and inflammation}\\ 
    \textbf{Summary}: \textit{Calcinosis, Cardiomegaly, Costophrenic Angle, Density, Nodule, Opacity, Pleural Effusion, Pulmonary Congestion, Pulmonary Edema, Thickening}\\
    \textbf{MTI annotated with Non-MEDLINE option}: \textit{Male humans pulmonary edema diagnosis, differential lung pleural effusion cardiomegaly inflammation infection tomography, x-ray computed retrospective studies}\\
    \textbf{Batch MTI indexer tool with Mesh On Demand}: \textit{Histamine pulmonary edema skin diagnosis, differential inflammation cardiomegaly pleural effusion} \\
    \textbf{NLU}: \textit{Normal}\\
    \textbf{NER}: \textit{Cardiomegaly, nodule} \\ \\
    
    \item \textbf{Findings}:  \textit{Heart size within normal limits. There is focal left lateral base airspace disease. There is a 6 mm nodular opacity in the right midlung. No pneumothorax. No pleural effusion. No displaced rib fractures. There is an apparent deformity of the right. humeral surgical neck. This is not seen on the comparison. Correlate clinically with history of fracture.} \\
    \textbf{Impression}: \textit{Left base airspace disease and nodular opacity in the right midlung.} \\
    \textbf{Summary}: \textit{Airspace Disease, Deformity, Opacity} \\
    \textbf{MTI annotated with Non-MEDLINE option}: \textit{Rib fractures pleural effusion exudates and transudates pneumothorax epiphyses humerus}\\
    \textbf{Batch MTI indexer tool with Mesh On Demand}: \textit{Diphosphoglyceric acids hemoglobins carbon dioxide oxygen hydrogen-ion concentration} \\
    \textbf{NLU}: \textit{6 mm}\\
    \textbf{NER}: \textit{Opacity, deformity} \\ \\
    
    \item \textbf{Findings}: \textit{The lungs and pleural spaces show no acute abnormality. XXXX scar in the right lateral midlung. Adjacent focal pleural thickening is noted. Chronic blunting of both lateral costophrenic XXXX. Heart size and pulmonary vascularity within normal limits. Tortuous, ectatic thoracic aorta, unchanged. XXXX sternotomy XXXX intact.} \\ 
    \textbf{Impression}:  \textit{No acute pulmonary abnormality.} \\ 
    \textbf{Summary}: \textit{Aorta, Thoracic,Cicatrix,Costophrenic Angle,Thickening}\\
    \textbf{MTI annotated with Non-MEDLINE option}: \textit{Humans respiratory system abnormalities pleural diseases lung pleural effusion pleural cavity tomography, x-ray computed thorax}\\
    \textbf{Batch MTI indexer tool with Mesh On Demand}: \textit{Cytarabine} \\
    \textbf{NLU}: \textit{Thoracic aorta}\\
    \textbf{NER}: \textit{Thickening} \\ \\
    
\end{enumerate}
    
\end{document}